\documentclass[twocolumn,showpacs,pra]{revtex4}

\usepackage{graphicx}
\usepackage{amsmath}

\begin{document}

\hyphenation{Feshbach Fesh-bach}

\title{ Feshbach resonances with large background scattering length: interplay with open-channel resonances }

\begin{abstract}
Feshbach resonances are commonly described by a single-resonance Feshbach model, and open-channel resonances are not taken into account explicitly. However, an open-channel resonance near threshold limits the range of validity of this model. Such a situation exists when the background scattering length is much larger than the range of the interatomic potential. The open-channel resonance introduces strong threshold effects not included in the single-resonance description. We derive an easy-to-use analytical model that takes into account both the Feshbach resonance and the open-channel resonance. We apply our model to $^{85}$Rb, which has a large background scattering length, and show that the agreement with coupled-channels calculations is excellent. The model can be readily applied to other atomic systems with a large background scattering length, such as $^6$Li and $^{133}$Cs. Our approach provides full insight into the underlying physics of the interplay between open-channel (or potential) resonances and Feshbach resonances.
\end{abstract}

\author{B.~Marcelis$^1$}
\author{E.~G.~M.~van~Kempen$^1$}
\author{B.~J.~Verhaar$^1$}
\author{S.~J.~J.~M.~F.~Kokkelmans$^{1,2}$}
\affiliation{$^1$Eindhoven University of Technology, P.O.~Box~513, 5600~MB  Eindhoven, The Netherlands \\
$^2$Laboratoire Kastler Brossel, Ecole Normale Sup\'erieure, 24 rue Lhomond, 75231 Paris 05, France}

\date{February 10, 2004}


\maketitle

\section{Introduction}

A magnetically induced Feshbach resonance~\cite{Feshbach58,Tiesinga92} is an indispensable tool to control the atom-atom interaction in ultracold gases. By simply changing the magnetic field around resonance, the s-wave scattering length, which is a measure for the strength of the interactions, can be given basically any value. Exactly on resonance, the scattering length is infinite, and its value is therefore much larger than any other lengthscale that characterizes the atomic gas system. This means that the scattering length effectively drops out of the physical problem. For very low temperatures and on resonance, universal behavior has been predicted~\cite{heiselberg,carlson}. For higher temperatures, or for situations further away from resonance, knowledge about the correct energy-dependence of the scattering phase shift will be needed to account for an accurate description of a resonant two-body interaction~\cite{ho,kokkelmans}.

Crossing the Feshbach resonance, the scattering length $a$ goes from positive to negative values through infinity, and the effective interaction changes accordingly from repulsive ($a>0$) to attractive ($a<0$). Using two atomic spin states of a fermionic gas, it is possible to study the crossover region between a BCS superfluid and a Bose-Einstein condensate (BEC) of molecules~\cite{ohashi,milstein,regal}.

The Feshbach resonance is associated with a molecular state just below the collision threshold. When the magnetic field is changed during an experiment, (quasi-bound) molecules can be formed~\cite{stenger,abeelen,mies} while sweeping through the resonance. At JILA, coherent oscillations between atoms and molecules have been observed by applying two magnetic field `Ramsey' pulses close to resonance~\cite{Donley02,Claussen03}. The oscillation frequency was directly related to the binding energy of the molecules. More recently, Feshbach resonances were utilized in the formation of ultracold molecules from a degenerate atomic Fermi gas~\cite{Regal03,Strecker03,Jochim03,Cubizolles03}. Here molecules could be produced reversibly by sweeping twice through resonance. In some cases these dimers have been cooled further down below the BEC transition temperature~\cite{Jochim03vrs2,Zwierlein03,Greiner03}. In the case of bosonic systems, Feshbach resonances have been utilized to produce ultracold sodium, cesium and rubidium molecules~\cite{Xu03,Chin03,Durr04}.

Close to the magnetic field value of resonance $B_0$, the s-wave scattering length $a$ shows a characteristic dispersive behavior and is given by
\begin{equation}
a(B) = a_\mathrm{bg} \left(1 - \frac{\Delta B}{B-B_0} \right).
\label{aofB}
\end{equation}
The background part of the scattering length $a_\mathrm{bg}$ summarizes the effect of the direct scattering processes in the open channel, \textit{without} coupling to other closed spin channels. $\Delta B$ is the field width of the Feshbach resonance. An example of this behavior can be seen in Fig.~\ref{aofB}. Since the Feshbach resonance is responsible for the dramatic behavior of the scattering length, the background process is usually considered not very interesting, and is in some cases even neglected. The direct process is then associated with non-resonant scattering, where the value of $a_{\rm bg}$ should be of order of the van der Waals potential range $r_0 \equiv (\mu C_6/8 \hbar^2)^{1/4}$~\cite{gribakin}, with $\mu$ the reduced mass and $C_6$ the van der Waals coefficient. This, however, is only true when there is no potential resonance close to the collision threshold.

\begin{figure}
\includegraphics[width=\columnwidth]{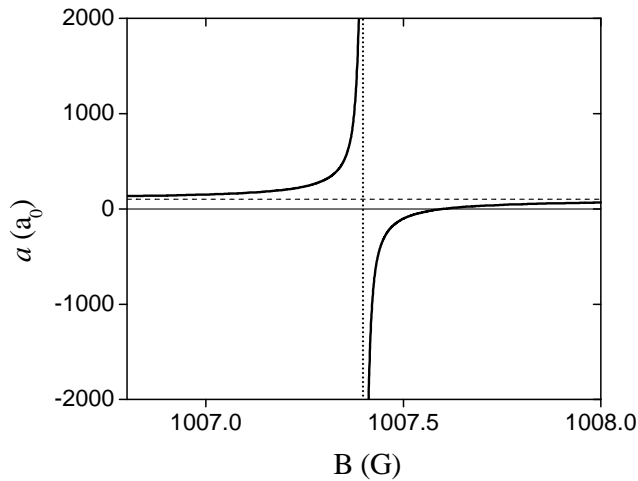}
\caption{Scattering length as a function of the magnetic field for $^{87}$Rb in the $|f,m_f\rangle = |1,1\rangle$ hyperfine channel. The horizontal dashed line indicates the direct scattering length $a_{\mathrm{bg}} = 100 \, \mathrm{a_0}$ (with $\mathrm{a_0}$ the Bohr radius), and the vertical dotted line indicates the resonant magnetic field $B_0 = 1007.4 \, \mathrm{G}$. The field width is given by $\Delta B = 0.2 \, \mathrm{G}$.} \label{scatlength}
\end{figure}

The effect of potential resonances can be easily estimated by comparing the value of $a_\mathrm{bg}$ to the potential range $r_0$. For example, in the case of $^{87}$Rb, the direct interaction potential is non-resonant since $a_{\rm bg}=100$~a$_0$~\cite{Kempen02} and $r_0$ is of the same order for rubidium. There are also situations where this is not the case. When $a_{\rm bg}$ is large and positive, there is an open-channel bound state very close to threshold. For example, this occurs for $^{133}$Cs, where $a_{\rm bg}=905$~a$_0$~\cite{Leo98,Julienne03} is much larger than $r_0$. In other situations $a_{\rm bg}$ can be large and negative, indicating that the interaction potential has a virtual state near threshold. Examples here are $^{85}$Rb with $a_{\rm bg}=-443$~a$_0$~\cite{Claussen03}, and $^6$Li where $a_{\rm bg}\approx -2000$~a$_0$~\cite{Abeelen97,Abraham97}. In these cases, the non-resonant description of the background scattering process is not correct. This can be seen from a comparison between the single-resonance Feshbach model and a full coupled-channels calculation over the range of energies of interest for current experiments, as will be shown in the following.

In this paper, we demonstrate how the potential resonance can be taken into account analytically, by properly describing the interplay between the open-channel resonance and the Feshbach resonance. The open-channel resonance is associated with a pole of the direct part of the scattering matrix. In our approach, we include this open-channel pole into the multi-channel Feshbach formalism. 
The resonance phenomena discussed here in the context of cold atomic collisions are quite general, and have been studied before in the context of nuclear physics \cite{othernuclear}, electron-atom collisions \cite{otherelectron}, and electron-molecule collisions \cite{Domcke83}. In previous work the potential resonance is indirectly included in the closed-channel subspace, however, the potential resonance is actually located in the open-channel subspace. In our approach, the distinction between open-channel and closed-channel resonances is clearly made, which results in a better understanding of the underlying physics.

The paper is outlined as follows. In Sec.~\ref{SecSimple} we introduce the basic ideas behind our approach. In Sec.~\ref{SecFeshbach} we briefly discuss the theory of Feshbach resonances, based on a projection-operator approach. In Sec.~\ref{SecGamow} we discuss the properties of potential resonances, and explain how the so-called Mittag-Leffler expansion can be used to account for these resonances. The Mittag-Leffler expansion is used in Sec.~\ref{SecInterplay} to discuss the interplay between Feshbach resonances and potential resonances. In Sec.~\ref{SecNumerical} we give a thorough discussion of the results of our model. We compare our model with other commonly used approaches in Sec.~\ref{SecOther}, and conclusions are drawn in Sec.~\ref{SecConclusions}.

\section{Basic ideas}\label{SecSimple}

In this section, we introduce the basic ideas which are needed to describe the interplay between potential resonances and Feshbach resonances. We first introduce the Feshbach resonance with a non-resonant background scattering process, i.e. with a background scattering length on the order of the range of the interatomic potential. Then we discuss a single-channel s-wave resonance, and its relation to the poles of the scattering matrix. In case the background scattering length is large, this indicates that the open channel is nearly resonant, even without coupling to other channels. Then, we investigate the effect of the open-channel resonance on the Feshbach resonance. We study two different cases: one with large positive and one with large negative $a_{\rm bg}$.

\subsection*{Feshbach resonances}

\begin{figure}
\includegraphics[width=\columnwidth]{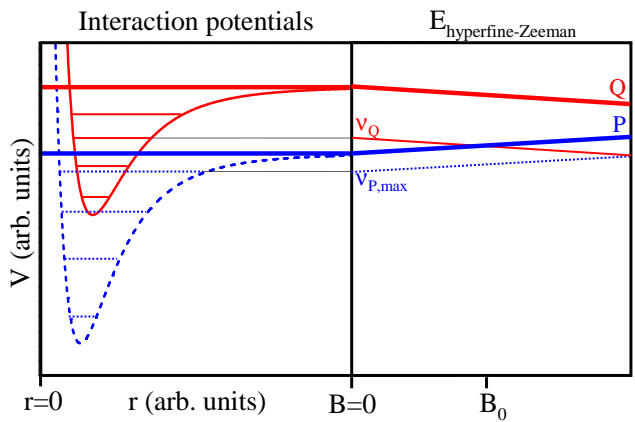}
\caption{(Color online) Left part of figure: Schematic illustration of the $P$-channel (dashed) and $Q$-channel (solid) interaction potential for $B=0$. Right part of figure: The interaction potentials are asymptotically connected to the magnetic field dependent two-particle hyperfine-Zeeman eigenenergies. Also shown are the highest $P$-channel vibrational bound state ($\nu_{P,\mathrm{max}}$) and a $Q$-channel vibrational bound state ($\nu_Q$) close to the collision threshold. At the resonant magnetic field $B_0$ the coupling between the $Q$-channel bound state and the $P$-channel scattering state becomes resonant.} \label{Potentials}
\end{figure}

The location of the vibrational bound states in the interaction potential determines the scattering properties in ultracold atomic collisions. In a multi-channel collision the incoming channel may be coupled to other (open and closed) channels during the collision. A channel is energetically open (closed) when the total energy of the two-atom system is above (below) the dissociation threshold energy of the incoming channel. In the Feshbach formalism the total Hilbert space describing the spatial and spin degrees of freedom is divided into two subspaces, indicated by $\mathcal{P}$ and $\mathcal{Q}$. Generally, the open channels are located in $\mathcal{P}$-space, and the closed channels in $\mathcal{Q}$-space. As a model example, we consider the case with only one open ($P$), and one closed ($Q$) channel. A schematic illustration of the interaction potentials associated with these two channels is shown in the left part of Fig.~\ref{Potentials}. Since the total energy is above the $P$-threshold, this $P$-channel is automatically the incoming and outgoing channel. Later on, we also consider the case where the total energy is below the $P$-threshold. Strictly speaking, the $P$-channel is then energetically closed and should belong to the $\mathcal{Q}$ subspace. However, for convenience we will still use the labelling $P$ and $Q$ to distinguish between these two channels.

In the case of alkali atoms, the $P$ and $Q$ channels are associated with different spin configurations, and have a different effective magnetic moment. The energy difference between the channels can be tuned by changing the external magnetic field. This means that the $P$ and $Q$ potentials can be shifted relative to each other. The Zeeman energy shift of the $P$-channel and $Q$-channel thresholds is schematically shown in the right part of the figure. Also indicated is the energy of the highest $P$-channel vibrational bound state ($\nu_{P,\mathrm{max}}$) and a $Q$-channel vibrational bound state ($\nu_Q$) close to the $P$-threshold.

During the collision the atoms are coupled from the $P$-channel to the $Q$-channel. The scattering process becomes resonant when a $Q$-channel bound state is located close to the $P$-channel collision threshold, giving rise to a Feshbach resonance. The unperturbed $Q$-channel bound state is dressed by the coupling to the $P$-channel. This dressed state can be considered as a \mbox{(quasi-)}bound state of the total scattering system. The colliding atoms are temporarily captured in the \mbox{(quasi-)}bound state, and after a characteristic time $\tau = 2 \hbar / \Gamma$, with $\Gamma$ the decay width of the \mbox{(quasi-)}bound state, they return to the $P$-channel. At the magnetic field value $B_0$, where the dressed state crosses the $P$-threshold, the scattering length has a singularity. The resulting behavior of the scattering length was already shown in Fig.~\ref{scatlength}.

The scattering length is related to the zero-energy limit of the scattering matrix. In the Feshbach description the energy-dependent scattering matrix is divided into two parts. The direct part $S^P_{\rm direct}$ describes the scattering processes in the $P$-channel, \textit{without} coupling to the $Q$-channel. The Feshbach resonance is described by the resonant part $S^Q_{\rm res}$. The resulting total scattering matrix $S=S^P_{\rm direct} S^Q_{\rm res}$ summarizes the energy and field dependence of the scattering process around resonance.

The direct part of the scattering process is usually assumed to be non-resonant. In the treatment by Moerdijk et al.~\cite{Moerdijk} the direct part of the scattering matrix is described by a single parameter only: $S^P_{\rm direct}(k)=\exp[-2ika_{\mathrm{bg}}]$, where $k$ is the relative wavenumber of the two colliding particles. In the case $a_{\mathrm{bg}}$ is of order of the range of the potential $r_0$, the resulting single-resonance Feshbach model accurately summarizes the scattering process for the energies and magnetic fields of interest.

\subsection*{Potential resonances}

The direct scattering properties are determined by the location of the bound states in the $P$-channel potential. In case the $P$-channel has a bound state close to threshold, this channel is nearly resonant and $a_{\rm bg}$ is large and positive. This has several important implications, which have not been considered before in the description of Feshbach resonances in cold atomic collisions. The $P$-channel potential resonance introduces a resonance energy dependence of $S^P_{\rm direct}$, not included in the simple description based on $a_{\rm bg}$ only. It is not straightforward to describe this resonance by means of a projection formalism analogous to the Feshbach approach, since the bound state is located in the open channel. Still, it is possible to understand the resonance behavior by studying the poles of the scattering matrix.

\begin{figure}
\includegraphics[width=5cm]{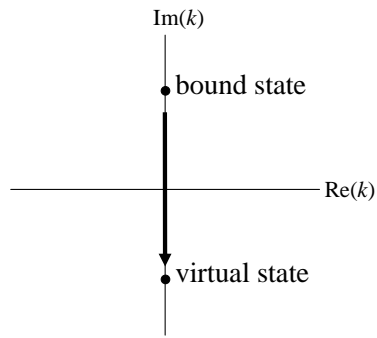}
\caption{Schematic illustration of the poles of the $S$ matrix in the complex $k$ plane. The bound-state pole is located at the positive imaginary $k$ axis. For a less attractive potential, this pole will move down the imaginary $k$-axis, giving rise to a virtual state.} \label{Poles_intro}
\end{figure}

There is a general connection between the resonances and bound states of a scattering system, and the various poles of the scattering matrix \cite{Taylor72}. For single-channel s-wave collisions, the poles in the upper half of the complex $k$-plane can only be located on the imaginary $k$-axis. Sufficiently close to the origin ($k=0$), the poles in the lower halfplane can only be located on the imaginary axis as well \cite{notepoles}. As a result, the energy $E=\hbar^2 k^2/2\mu$ associated with these poles will be strictly negative. The $P$-channel bound state is related to a pole of $S^P_{\rm direct}$ on the positive imaginary $k$-axis, as indicated in Fig.~\ref{Poles_intro}. If the bound state is located just below threshold, this pole is located close to the origin.

Now imagine that the potential gradually becomes less attractive. Then, the bound states will move towards threshold, and the poles of the $S$ matrix will move down the imaginary $k$-axis. Consequently, the bound-state pole close to the origin will move towards $k=0$, cross the origin when the bound state is located exactly at threshold, and eventually turn into a so-called virtual-state pole on the negative imaginary $k$-axis. The energy associated with this virtual state is negative, but there is no proper physical bound state associated with this energy. A virtual state can be regarded as a nearly-bound state that behaves much like a real bound state in the inner region of the interaction potential. Only in the asymptotic region ($r\rightarrow \infty$) the virtual state `discovers' it does not quite fit to the size of the interaction potential, and the virtual state exponentially explodes.

The direct part of the scattering matrix can be divided into a resonant and a non-resonant part: $S^P_{\rm direct} = S^P_{\rm bg} S^P_{\rm res}$. The background part is now related to the true range of the potential, while the resonant part takes the form:
\begin{equation}
S^P_{\rm res}(k) = -\frac{k+i\kappa}{k-i\kappa},
\end{equation}
where $i\kappa$ is the location of the pole on the imaginary $k$-axis. In the zero-energy limit, the direct part of the scattering matrix is related to the background scattering length as $a_{\rm bg} = a_{\rm bg}^P + 1/\kappa$. Here $a_{\rm bg}^P$ is of order $r_0$, while the resonant contribution of the $P$-channel is given by $1/\kappa$. The important effect of a bound or virtual state near threshold is the introduction of a resonance energy dependence of $S^P_{\rm direct}$. More details will be given in Sec.~\ref{SecGamow}.

\subsection*{Interplay}

Now we return to a multi-channel collision, with a Feshbach resonance resulting from coupling to $\mathcal{Q}$-space, and a potential resonance associated with $\mathcal{P}$-space. The total scattering matrix takes the form $S=S^P_{\rm bg} S^P_{\rm res} S^Q_{\rm res}$, where the background part is related to the true range of the potential. In a multi-channel problem, the bound and quasi-bound states of the scattering system are related to the poles of the total $S$ matrix. The bound-state poles are located on the positive imaginary $k$-axis, whereas the quasi-bound state poles are located in the lower half of the complex $k$-plane. These poles are generally not located on the imaginary axis, as they are related to complex energies $E_R - i\Gamma/2$. Here $E_R$ is the energy, and $\Gamma$ the decay width of the quasi-bound state. The decay width is related to the lifetime of the quasi-bound Feshbach resonance state.

\begin{figure}
\includegraphics[width=\columnwidth]{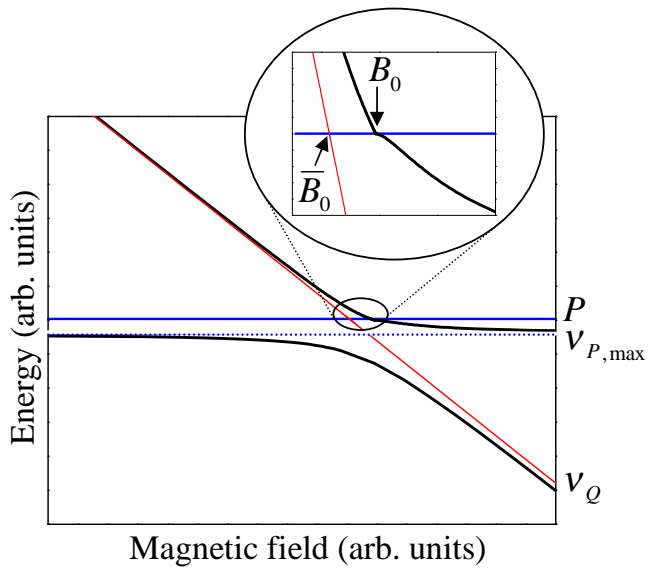}
\caption{(Color online) Example of an avoided crossing between a $P$-channel and a $Q$-channel bound state. The energy of the bare bound states is indicated by the horizontal and slanted lines. The energies of the dressed states are indicated by the solid lines. One of the dressed states crosses the $P$-threshold at $B_0$. The bare $Q$-channel bound state crosses the $P$-threshold at $\overline{B}_0$. Note that the energies are give relative to the $P$-channel collision threshold (horizontal solid line).} \label{Avoided_bound}
\end{figure}

\begin{figure}
\includegraphics[width=\columnwidth]{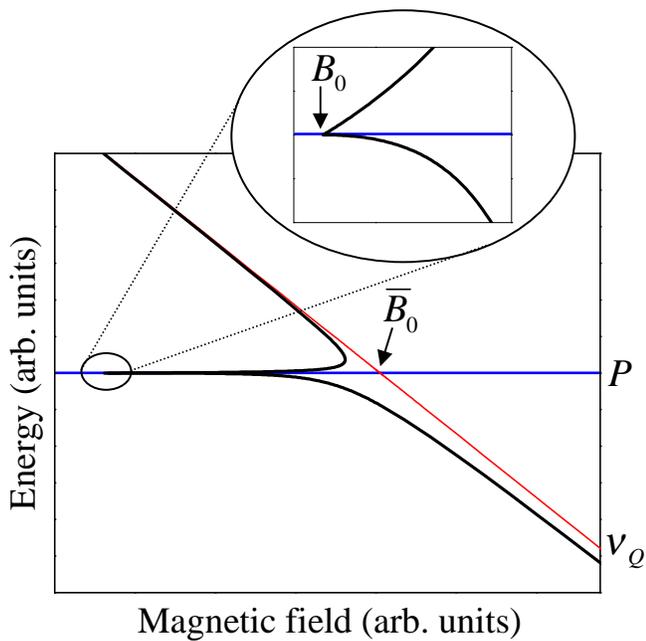}
\caption{(Color online) Effect of virtual state on Feshbach resonance in $^{85}$Rb in the $|2,-2\rangle$ hyperfine channel. Shown are the energy of the unperturbed $Q$-channel bound state (dotted line), and of the \mbox{(quasi-)}bound state (solid line) which is `dressed' by the coupling to the $P$-channel. The \mbox{(quasi-)}bound state crosses the collision threshold at $B_0$, the unperturbed $Q$-channel bound state crosses the collision threshold at $\overline{B}_0$. Note that the energies are give relative to the $P$-channel collision threshold (horizontal solid line).} \label{Avoided_virtual}
\end{figure}

In Fig.~\ref{Avoided_bound}, we consider the case where the $P$-channel has a bound state just below threshold. This means that $a_{\rm bg}$ will be large and positive, which is the case for $^{133}$Cs, for example. The bare bound states of the $P$ and $Q$ channels are indicated by the horizontal ($\nu_{P,{\rm max}}$) and slanted ($\nu_Q$) lines, respectively. These bare states are eigenstates in the uncoupled $\mathcal{P}$ and $\mathcal Q$ subspaces. The bare $Q$-channel bound state crosses the $P$-channel threshold at the magnetic field value $\overline{B}_0$. The dressed states are indicated by the solid lines.

Below the $P$-threshold, the behavior of the dressed states resembles an avoided crossing as described by a two-level Landau-Zener model. In the inset, the behavior of the dressed state is shown close to the $P$-channel threshold. At $B_0$ one of the dressed bound states crosses the collision threshold, acquires a decay width, and turns into a quasi-bound state. Exactly at $B_0$, the s-wave scattering length goes through infinity. Note that the binding energy of the dressed state curves quadratically towards threshold according to the well-known relation $\epsilon_{\rm bind}(B)=-\hbar^2/(2\mu a^2(B))\propto (B-B_0)^2$, as follows from Eq.~(\ref{aofB}). At threshold, the slope of the dressed state energy shows a discontinuity resulting from a non-zero decay width above threshold. We stress that, although the behavior of the dressed states shows some resemblance with the two-level Landau-Zener description, this model does not include these threshold effects and can not be used to properly describe the interplay between a potential resonance and a Feshbach resonance.

For $a_{\rm bg}$ large and negative, there is a $P$-channel virtual state near threshold. Examples here are $^{85}$Rb and $^6$Li. An example of the energy of the dressed states for this situation is given in Fig.~\ref{Avoided_virtual}. Here the difference with the Landau-Zener model is even more striking. The behavior of the dressed state energies can not be understood on the basis of an avoided crossing between the $Q$-channel bound state and the $P$-channel threshold. Based on a Landau-Zener approach, the dressed state above $P$-threshold would be located at the righthand side of the bare $Q$-channel bound state, which clearly is not the case. The binding energy still curves quadratically towards threshold, which it crosses at the magnetic field $B_0$. From the inset it can be seen that the slope of the dressed state energy shows a discontinuity at threshold, which is even more distinct than the situation given in Fig.~\ref{Avoided_bound}. For larger collision energies, the dressed quasi-bound state curves back to the bare $Q$-channel bound state \cite{notedipolar}.

In the following we derive an analytical model that describes the interplay between potential resonances and Feshbach resonances. We used this model to obtain Figs.~\ref{Avoided_bound} and \ref{Avoided_virtual}, which are model examples applicable to $^{133}$Cs and $^{85}$Rb, respectively. In Sec.~\ref{SecNumerical} we show that the agreement between our model and full coupled-channels calculations is excellent for a large domain of energies and magnetic fields, while the single-resonance Feshbach model gives poor results in the same domain. We compare our model with other approaches in Sec.~\ref{SecOther}.

\section{Feshbach Resonance Theory}\label{SecFeshbach}
Feshbach resonances in two-body collisions are related to the coupling of different spin channels, and can be conveniently described in an approach due to Feshbach \cite{Feshbach58,Feshbachboek}. In this approach the total Hilbert space $\mathcal{H}$ describing the spatial and spin degrees of freedom is divided into two subspaces $\mathcal{P}$ and $\mathcal{Q}$. In general $\mathcal{P}$ contains the open channels and $\mathcal{Q}$ the closed channels. The $S$ and $T$ matrices, which are related to the transition probabilities of the scattering process, are separated in two parts accordingly. The $P$-part describes the direct interactions in the open-channel subspace, and the $Q$-part describes the effect of the coupling to the closed-channels. Usually the $Q$-part contains the resonances, and the $P$-part is assumed to be non-resonant. We discuss in Sec.~\ref{SecGamow} why in some cases the $P$-part can introduce resonant features as well. In Sec.~\ref{SecInterplay} we show how these $P$-channel resonances can be taken into account analytically. First we discuss the Feshbach projection formalism in this section.

One can construct projection operators $P$ and $Q$, which project onto the subspaces $\mathcal{P}$ and $\mathcal{Q}$, respectively. The Schr\"{o}dinger equation for the two-body collision can then be written as a set of coupled equations: \begin{eqnarray} ( E_{\mathrm{tot}} - H_{PP} ) |\Psi_P \rangle & = & H_{PQ} |\Psi_Q \rangle , \label{hpp} \\ ( E_{\mathrm{tot}} - H_{QQ} ) |\Psi_Q \rangle & = & H_{QP} |\Psi_P \rangle .\label{hqq} \end{eqnarray} Here we use the notation $|\Psi_P \rangle \equiv P|\Psi \rangle$, $|\Psi_Q \rangle \equiv Q|\Psi \rangle$, $H_{PP} \equiv PHP$, $H_{PQ} \equiv PHQ$, etc., and $E_{\mathrm{tot}}=E+E_{\mathrm{thr}}$ is the total energy of the colliding atoms, with $E$ the kinetic energy and $E_{\mathrm{thr}}$ the energy of the open-channel threshold. As already mentioned, we are only interested in scattering processes with only one open channel. The $P$-channel is then simultaneously the incoming and outgoing channel. Also note that all energies in the following are given with respect to the open-channel collision threshold.

We multiply Eq.~(\ref{hqq}) from the left with the resolvent (or Green's) operator $G_{QQ}(E^+) \equiv [E^+ - H_{QQ}]^{-1}$: \begin{equation} |\Psi_Q \rangle = \frac{1}{E^+ - H_{QQ}}H_{QP} |\Psi_P \rangle, \end{equation} where $E^+=E+i\delta$ with $\delta$ approaching zero from positive values. Substituting the expression for $|\Psi_Q \rangle$ in Eq.~(\ref{hpp}), the problem in the $\mathcal{P}$ subspace is equivalent to solving the Schr\"{o}dinger equation $( E - H_{\mathrm{eff}} ) |\Psi_P \rangle = 0,$ where the effective Hamiltonian is given by \begin{equation} \label{effham} H_{\mathrm{eff}} = H_{PP} + H_{PQ} \frac{1}{E^+ - H_{QQ}} H_{QP}. \end{equation} The first term in the effective Hamiltonian describes the direct effect of the open-channel subspace $\mathcal{P}$ on the scattering process. The second term in the effective Hamiltonian describes the coupling of $\mathcal{P}$ space to $\mathcal{Q}$ space, propagation through $\mathcal{Q}$ space, and coupling back to $\mathcal{P}$ space again.

The resolvent operator $G_{QQ}$ can be expanded in terms of the discrete ($|\phi_i \rangle$) and continuum ($|\phi(\epsilon)\rangle$) eigenstates of $H_{QQ}$: \begin{equation}  \frac{1}{E^+ - H_{QQ}} = \sum_i \frac{|\phi_i \rangle\langle\phi_i |}{E - \epsilon_i^Q } + \int \frac{|\phi(\epsilon) \rangle\langle \phi(\epsilon)|}{E^+ - \epsilon} \mathrm{d}\epsilon. \end{equation} Now suppose there is only one discrete bound state of $H_{QQ}$ for which the eigenvalue $\epsilon_b^Q$ is close to the collision energy $E$ of the scattering state. That is, the other $Q$-channel bound states are located sufficiently far away from the $P$-threshold so that their contribution can be safely neglected. In this case we can also neglect the continuum expansion of $H_{QQ}$, and the problem in the $\mathcal{P}$ subspace reduces to \begin{equation} ( E - H_{PP} ) |\Psi_P \rangle= H_{PQ} \frac{|\phi_b \rangle\langle\phi_b|}{E - \epsilon_b^Q}H_{QP}|\Psi_P \rangle. \end{equation}

Now we can formally solve the coupled problem by multiplying from the left with the resolvent operator $G_{PP}(E^+) \equiv [E^+ - H_{PP}]^{-1}$: \begin{equation} \label{formalP} |\Psi_P \rangle= |\Psi^+_P \rangle+ \frac{1}{E^+ - H_{PP}}H_{PQ} \frac{ |\phi_b \rangle\langle \phi_b |}{E - \epsilon_b^Q } H_{QP} |\Psi_P \rangle. \end{equation} The scattering state $|\Psi^+_P \rangle$ corresponds to the homogeneous part of the Schr\"{o}dinger equation projected into $\mathcal{P}$ space, and is an eigensolution of $H_{PP}$ with outgoing spherical wave boundary conditions. The scattering states are energy-normalized as $\langle \Psi_P^+ (E)| \Psi_P^+ (E')\rangle = \delta(E-E')$, where $\delta(E)$ is the Dirac delta function.

Before calculating the $S$ and $T$ matrices, we introduce the Lipmann-Schwinger equations for $|\Psi^{\pm}_P \rangle$, where the superscript + (-) indicates outgoing (incoming) spherical wave boundary conditions \cite{Joachain72}: \begin{equation} |\Psi^{\pm}_P \rangle = |\chi_P \rangle + \frac{1}{E^{\pm} - H_{PP}}V_{PP}|\chi_P \rangle. \label{psiplusmin} \end{equation} Here the Hamiltonian in $\mathcal{P}$ space is written as $H_{PP} \equiv H^0_{PP} + V_{PP}$, where $H^0_{PP}$ represents the sum of the relative kinetic energy operator and the two-particle hyperfine-Zeeman interactions, and $V_{PP}$ the interatomic interactions, both projected onto the $\mathcal{P}$-channel subspace. The unscattered states $|\chi_P \rangle$ are eigenstates of $H^0_{PP}$.

The $T_P$ matrix, giving the transition amplitude due to scattering in the $\mathcal{P}$ subspace only, can now be calculated as \begin{eqnarray} \langle \chi_P|\mathcal{T}_P|\chi_P \rangle & \equiv & \langle \chi_P|V_{PP} \left[ 1 + \frac{1}{E^+ - H_{PP}}V_{PP} \right] |\chi_P \rangle \nonumber\\ & = & \langle \chi_P|V_{PP} |\Psi^+_P \rangle. \end{eqnarray} Introducing the second term in the effective Hamiltonian Eq.~(\ref{effham}), according to the two-potential theorem \cite{Joachain72} the total $T$ matrix is given by \begin{eqnarray} \langle \chi_P|\mathcal{T}|\chi_P \rangle & = & \langle \chi_P|\left[ V_{PP}+H_{PQ}\frac{1}{E^+ - H_{QQ}}H_{QP} \right]|\Psi_P \rangle \nonumber\\ & = & \langle \chi_P|V_{PP}|\Psi^+_P \rangle \nonumber\\ && + \langle \Psi^-_P|H_{PQ} \frac{1}{E^+ - H_{QQ}}H_{QP}|\Psi_P \rangle, \end{eqnarray} where we have used the formal solution for $|\Psi_P \rangle$ given by Eq.~(\ref{formalP}) together with relation Eq.~(\ref{psiplusmin}). We see that apart from the direct term $T_P$, the transition amplitude contains a term which results from the coupling to the closed-channel subspace $\mathcal{Q}$.

The relation between the $S$ matrix $S(k)=\langle \chi_P|\mathcal{S}|\chi_P \rangle$ and the $T$ matrix $T(k)=\langle \chi_P|\mathcal{T}|\chi_P \rangle$ is given by the operator equation $\mathcal{S}=1-2\pi i\mathcal{T}$. From this equality, together with the expansion of the resolvent $G_{QQ}$, it follows that the $S$ matrix of the effective problem in $\mathcal{P}$ space is given by \begin{equation} S(k) = S_P(k) - 2\pi i \frac{\langle \Psi^-_P|H_{PQ} |\phi_b \rangle  \langle \phi_b | H_{QP} |\Psi_P \rangle}{E - \epsilon_b^Q }. \end{equation}
Here $S_P(k)=\langle \Psi^-_P(E)|\Psi^+_P(E) \rangle$ is the direct part of the $S$ matrix, which describes the effect of the scattering process in $\mathcal{P}$ space only, without coupling to the closed channels in $\mathcal{Q}$ space.

We can finally solve for $|\Psi_P \rangle$ by multiplying Eq.~(\ref{formalP}) from the left with $\langle \phi_b|H_{QP}$, which results in
\begin{equation}
    S(k) =  S_P(k) \left(1 - 2\pi i \frac{\left| \langle \phi_b | H_{QP} |\Psi^+_P \rangle \right|^2}{E - \epsilon_b^Q - A(E)} \right). \label{smatrix}
\end{equation}
The term \begin{equation} A(E) \equiv \langle \phi_b|H_{QP} \frac{1}{E^+ - H_{PP}} H_{PQ}|\phi_b \rangle \label{AE} \end{equation} in the denominator is the complex energy shift, which is the energy difference between the bare bound state $|\phi_b \rangle$ and the dressed \mbox{(quasi-)}bound state.

Inserting a complete set of eigenstates of $H_{PP}$, the complex energy shift can be written as $A(E) = \Delta_{\mathrm{res}}(E)-\frac{i}{2}\Gamma(E)$. The real part $\Delta_{\mathrm{res}}(E)$ shifts the unperturbed energy $\epsilon_b^Q$, and the imaginary part $\Gamma(E)$ turns the unperturbed bound state $|\phi_b \rangle$ into a quasi-bound state. Note that $A(E)$ is purely real for energies below the $P$-threshold (i.e., $E<0$). The energy of the dressed states can be found by solving for the poles of the $S$ matrix.

In the low-energy domain it is usually assumed that the real part of the energy shift can be taken as approximately constant, $\Delta_{\mathrm{res}}(E)\simeq \Delta_{\mathrm{res}}(E=0)$. The energy width $\Gamma(E)=2\pi |\langle \phi_b | H_{QP} |\Psi^+_P(E) \rangle|^2$ is proportional to $k$ in the limit for $k\downarrow 0$, which is a consequence of the surviving s-wave part of the scattering wave function. Therefore it is usually assumed that $\Gamma(E) \simeq 2Ck$, with $C$ a constant that characterizes the coupling strength between $P$ and $Q$ \cite{Moerdijk}. The resulting expression for the $S$ matrix \begin{equation} S(k) = S_P(k) \left( 1-\frac{2iCk}{E - \epsilon_b^Q(B) - \Delta_{\mathrm{res}} + iCk} \right) \label{singleresonance} \end{equation} will be referred to as the \textit{single-resonance} approximation.

The direct part of the $S$ matrix is related to the direct part of the scattering length as $S_P (k) = \exp[-2ika_{\mathrm{bg}}]$, and the second term describes the resonant behavior due to coupling to $Q$-space. In the limit $k \downarrow 0$ this single-resonance expression for the $S$ matrix gives the dispersive behavior of Eq.~(\ref{aofB}).

However, the single-resonance approximation has a limited range of validity. More specifically, it is based on the assumption that the scattering wave function $|\Psi_P^+\rangle$ can be approximated by its low-energy s-wave limit, which is proportional to $k^{1/2}$ (also known as Wigner's threshold law). If the $P$-channel has a low-energy resonance, such as a (nearly-)bound state (with $\epsilon_{\rm res}^P = -\hbar^2 \kappa^2/2\mu$) close to threshold, this approximation breaks down when $|E/\epsilon_{\rm res}^P| \ll 1$ is not satisfied.

In the following we show that the assumptions about the trivial energy dependence of $S_P(k)$ and $A(E)$ are not valid in the case the $P$-channel is nearly resonant. We derive how the full energy dependence can be taken into account analytically. In Sec.~\ref{SecGamow} we discuss the Mittag-Leffler expansion of the resolvent operator $[E^+ -H_{PP}]^{-1}$ based on Gamow resonance states, which takes these $P$-channel resonances into account. In Sec.~\ref{SecInterplay} we derive analytical expressions for the energy shift and width based on this expansion.

\section{Open-Channel Resonances}\label{SecGamow}
Without coupling to the closed-channel subspace $\mathcal{Q}$, the scattering properties in the open-channel subspace $\mathcal{P}$ are governed by the uncoupled Schr\"{o}dinger equation $(E - H_{PP}) |\Psi_P \rangle = 0$, which has scattering solutions $|\Psi^+_P \rangle$. We assume that the interatomic potential is of finite range, i.e., it can be neglected beyond some radius $R_0$ and is of the form \begin{equation} V_{PP}(r) = \left\{ \begin{array}{ll} V_{PP}(r) & \qquad \textrm{for } 0 \leq r<R_0, \\ 0 & \qquad \textrm{for } r\geq R_0. \end{array} \right. \end{equation} The potentials used in the description of weakly interacting cold alkali gases have the asymptotic form $V(r) \sim -C_6/r^6$. If we take $R_0$ such that $V(R_0)$ is small compared to the asymptotic kinetic energy $E$, the finite-range approximation can be applied retaining high accuracy \cite{notefiniterange}.

The direct part of the $S$ matrix is determined by the uncoupled $\mathcal P$-space scattering solutions. In the low-energy limit the scattering properties for single-channel problems are determined by the s-wave part of the scattering wave function $\psi_P^{\pm}(\mathbf{r})=\langle \mathbf{r}|\psi^{\pm}_P \rangle$. If there are no low-energy potential resonances, the scattering length $a_{\mathrm{bg}}$ of the uncoupled $H_{PP}$ problem encapsulates the direct part of the scattering process in the low-energy domain.

However, if $H_{PP}$ possesses a low-energy potential resonance this is not true in general. The $S_P$ matrix contains a resonant feature and becomes strongly energy dependent near the collision threshold. The direct part of the $S$ matrix takes the form $S_P = S_{\mathrm{bg}}^P S_{\mathrm{res}}^P$, where the background part of the $S$ matrix, described by $S_{\mathrm{bg}}^P = \exp[-2ik \, a_{\mathrm{bg}}^P]$, summarizes the effect of the $P$-channel potential on the scattering solutions, \textit{excluding} the resonant part. The resonant part is described by $S_{\mathrm{res}}^P$. Note that we added a superscript $P$ to distinguish the $P$-channel resonance from the Feshbach resonance, which is induced by coupling to the $Q$-channel.

The resonances and bound states of the $P$-channel interaction potential correspond to the poles of $S_P$. More specifically, the energy dependence of the resonant part $S_{\mathrm{res}}^P$ is determined by the analytical properties of the ($P$-channel) Jost function $\mathcal{F}(k)$ \cite{Newton82}. The Jost function is related to $S_P$ matrix as $S_P(k) = \mathcal{F}(-k) / \mathcal{F}(k)$, and the zeros of the Jost function correspond to the poles of the $S_P$ matrix.

The Jost function $\mathcal{F}(k)$ is an entire function in $k$ for finite-range potentials. In the general case, the zeros of $\mathcal{F}(k)$ in the upper halfplane are only located at the imaginary $k$ axis, are finite in number, and correspond to the bound states in $\mathcal{P}$ space. Note that since $E<0$ in this case, the $\mathcal{P}$ subspace is in fact also energetically closed. There is a countable, infinite number of zeros of $\mathcal{F}(k)$ in the lower halfplane. The poles located in the third and fourth quadrant correspond to the resonances associated with a non-zero angular momentum of the colliding particles, and are located symmetrically with respect to the imaginary $k$ axis. The poles located at the negative imaginary axis are finite in number and correspond to the s-wave virtual states. Labelling the zeros by $k_n$, the Jost function can be written as the product~\cite{Newton82} \begin{equation} \mathcal{F}(k) = \prod_{n=1}^{\infty} \left( 1-\frac{k}{k_n} \right) h(k), \end{equation} where $h(k)$ is a smooth function related to the range of the potential $r_0$. This expression is normalized such that it reproduces the correct behavior near threshold, $\mathcal{F}(k) \rightarrow 1$ for $k \rightarrow 0$. Note that the physical parameter $r_0$ should not be confused with the cut-off radius $R_0$, and typically $r_0 \ll R_0$.

From the relation $E=\hbar^2 k^2/2\mu$ it follows that the mapping from $k$ to $E$ is two-to-one. As a result the single-valued functions of $k$, such as the Jost function and the $S$ matrix, will be double-valued functions of the complex energy $E$. However, we can map the upper and lower half of the complex $k$ plane to two different complex energy planes. The upper half ($\mathrm{Im}(k)>0$) is mapped onto the so-called physical (or first) sheet, whereas the lower half ($\mathrm{Im}(k)<0$) is mapped onto the non-physical (or second) sheet. If we consider the $E$ plane as a two-sheeted Riemann surface \cite{Byron92}, the afore mentioned functions are single-valued functions of the energy $E$. Both sheets have a branch-cut discontinuity on the real axis running from 0 to $\infty$. The two Riemann sheets are connected by the positive real $E$-axis, where $\mathrm{Im}(k)=0$.

The bound states and resonances of a scattering system correspond to the poles of $S(k)$. If we consider the scattering matrix as a function of the collision energy, which we will loosely write as $S(E)$, there is a similar relation between the poles of $S(E)$ on the two Riemann sheets and the bound and resonance states of a scattering system. The bound states correspond to the poles of $S(E)$ at the negative real $E$ axis on the physical (or first) Riemann sheet. The $l \neq 0$ resonances correspond to the poles of $S(E)$ in the first and fourth quadrant of the complex $E$ plane on the non-physical (or second) Riemann sheet. The s-wave virtual states correspond to the poles of $S(E)$ at the negative real $E$ axis on the non-physical Riemann sheet.

\subsection*{Virtual states}

\begin{figure}
\includegraphics[width=\columnwidth]{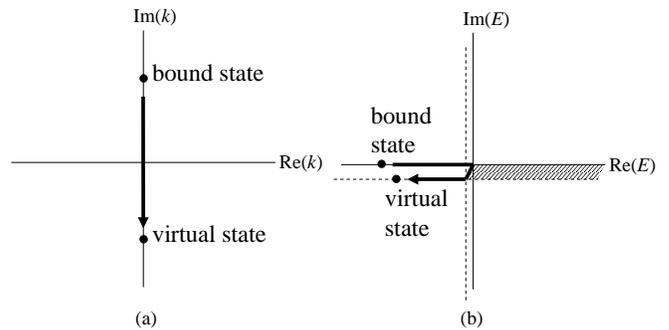}
\caption{Schematic illustration of the movement of the poles of the $S$ matrix in the complex $k$ plane (a) and the two-sheeted complex $E$ plane (b). The solid $E$ axes correspond to the first (or physical) Riemann sheet, while the dashed $E$ axes correspond to the second (or non-physical) Riemann sheet. Both sheets have a branch-cut discontinuity on the real axis running from 0 to $\infty$. The two Riemann sheets are connected by the positive real axis, where $\mathrm{Im}(k)=0$. Initially, the pole is located at the positive imaginary $k$ axis, which corresponds to a bound-state pole on the negative real $E$ axis on the first Riemann sheet. If the interaction potential effectively gets less attractive the pole moves towards $k=0$, at some point crosses the origin, and turns into a virtual-state pole at the negative imaginary $k$-axis. This corresponds to a virtual-state pole at the negative real $E$-axis on the second Riemann sheet.} \label{Poles}
\end{figure}

If the potential $V$ effectively gets less attractive, the bound states of the potential will move towards the collision threshold. At some point the highest bound state will cross the collision threshold, and turn into a virtual state in the case of s-wave collisions. The corresponding transition of a bound-state pole into a virtual-state pole is schematically illustrated in Fig.~\ref{Poles}. A virtual state behaves much like a bound state in the inner region of the potential, but it is not a proper bound state as it behaves asymptotically as $e^{+|\kappa| r}$. The important effect of a low-energy virtual state is that it introduces a resonance feature in the $S_P$ matrix.

In Sec.~\ref{SecNumerical} we calculate the position of the poles of the $S_P$ matrix in the case of $^{85}$Rb, and find a virtual state at $\epsilon_{\mathrm{virtual}}/k_B = - 6.45 \, \mu \mathrm{K}$. This is close to threshold, since the typical energies in current experimental setups are of order microKelvin. We discuss how this virtual state can be accounted for in the following. The zero of the Jost function associated with the virtual state is labelled by $-i \kappa_{\mathrm{vs}}$, where $\kappa_{\mathrm{vs}}$ is a positive constant. Furthermore, all other poles of the $S_P$ matrix are located far from the origin $k=0$, and the contribution of these distant poles to the $S_P$ matrix is summarized by a smooth function $g(k)$. The Jost function can then be written as \begin{equation} \mathcal{F}(k) = \left(1 + \frac{k}{i\kappa_{\mathrm{vs}}} \right)g(k). \end{equation} For real $k$ it follows that the direct part of the $S$ matrix is given by \begin{equation} S_P(k) = e^{-2ika_{\mathrm{bg}}^P} \frac{i\kappa_{\mathrm{vs}}-k}{i\kappa_{\mathrm{vs}}+k}, \end{equation} where the background factor $\exp[-2ika_{\mathrm{bg}}^P]$ summarizes the effect of all the non-resonant poles of the $S_P$ matrix.

The scattering phase $\delta(k)$ is related to the $S_P$ matrix as $S_P(k)=\exp[2i\delta(k)]$, and is evaluated as \begin{equation} \delta(k) = -ka_{\mathrm{bg}}^P + \arctan\left[ \frac{k}{\kappa_{\mathrm{vs}}} \right].  \label{deltavirtual} \end{equation} The background part is related to the phase of $g(k)$, and is linear in $k$ in the low-energy limit \cite{Note2}. The resonant part is related to the pole at $k=-i\kappa_{\mathrm{vs}}$, and causes a `bump' in the scattering phase at low energies. Moreover, if the virtual-state pole gets closer to threshold ($\kappa_{\mathrm{vs}} \rightarrow 0$), the scattering length $a_{\mathrm{bg}}=a_{\mathrm{bg}}^P - 1/\kappa_{\mathrm{vs}}$ will become more and more negative.

\begin{figure}
\includegraphics[width=\columnwidth]{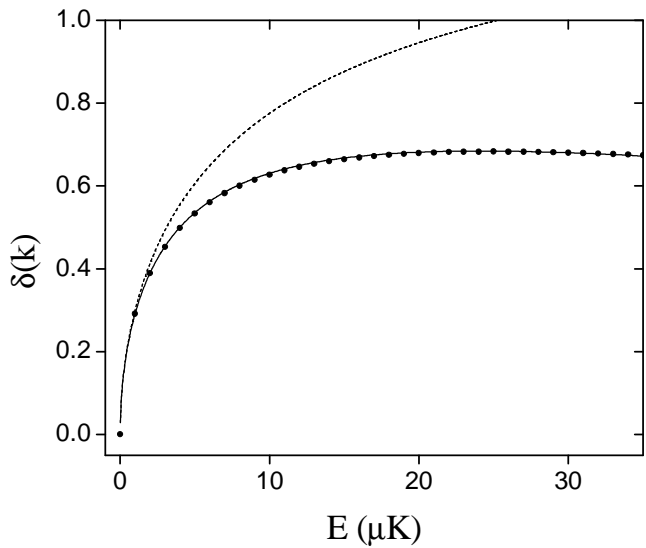}
\caption{Scattering phase $\delta(k)$ for $^{85}$Rb in the $|f,m_f\rangle = |2,-2 \rangle$ spin channel. The black dots represent the numerical results. The solid line is obtained from Eq.~(\ref{deltavirtual}). The dashed line is obtained from the contact potential approximation $\delta(k) = -\arctan[ka_{\mathrm{bg}}]$. The scattering length of $^{85}$Rb in this channel is $a_{\mathrm{bg}} = -443 \, \mathrm{a_0}$, in agreement with coupled-channels calculations. The background scattering length is $a_{\mathrm{bg}}^P = +119 \, \mathrm{a_0}$, and the virtual state is located at $k_{\mathrm{vs}}= -i\kappa_{\mathrm{vs}} = -1.78\cdot 10^{-3}i \, [1/\mathrm{a_0}]$, which corresponds to the term $-1/\kappa_{\mathrm{vs}} = -562 \, \mathrm{a_0}$ in the expression for $a_{\mathrm{bg}}$.} \label{Phase85Rb}
\end{figure}

To show that this virtual state has to be taken into account explicitly, Fig.~\ref{Phase85Rb} shows the scattering phase for $^{85}$Rb in the $|f,m_f\rangle = |2,-2 \rangle$ $P$-channel, without coupling to the $Q$-channels. The black dots represent the numerical results, which are obtained by solving the Schr\"{o}dinger equation using the proper physical and state-of-the-art rubidium potentials. The solid line is obtained from the virtual-state expression, and the dashed line is obtained from the usual contact potential approximation $\delta(k) = -\arctan[ka_{\rm bg}]$ \cite{Taylor72}. Let us stress again that Fig.~\ref{Phase85Rb} shows the scattering phase for the $P$-channel only, and the coupling to the $Q$-channel has been excluded from this calculation. For this particular channel $a_{\mathrm{bg}} = -443 \, \mathrm{a_0}$ \cite{Claussen03}. The virtual state contributes to the scattering length as $-1/\kappa_{\mathrm{vs}} = -562 \, \mathrm{a_0}$, and $a_{\mathrm{bg}}^P = +119 \, \mathrm{a_0}$ is now of the order of the potential range $r_0$. More details about the numerical calculation can be found in Sec.~\ref{SecNumerical}.

Comparing the virtual-state expression for $\delta(k)$ with the numerical results, the agreement is excellent. If we compare the numerical results with the contact potential expression, it is immediately seen that this expression already starts to deviate significantly at a few microKelvin. This indicates that the scattering length parameter only does not fully encapsulate the energy dependence of the scattering physics, and the $P$-channel resonance should be taken into account explicitly.

\subsection*{Mittag-Leffler series}

The $S$ matrix, $T$ matrix, and the resolvent (or Green's) operator $G_{PP}(E)\equiv [E-H_{PP}]^{-1}$ have their poles in common \cite{Note1}. This suggests it is possible to expand the resolvent in a Mittag-Leffler series~\cite{Newton82}, where the resolvent is written as a sum over the different pole contributions. The corresponding residues are related to the bound and resonance states, which are eigensolutions of $H_{PP}$. The scattering and transition matrix can then be written as a sum over the pole contributions as well.

Eigensolutions associated with poles of the scattering matrix were first introduced by Gamow in his theoretical description of $\alpha$-decay \cite{Gamow28}. In the case of s-wave scattering, the Gamow function $\Omega_n(r)=\langle \mathbf{r}|\Omega_n \rangle$ associated with the pole $k_n$ of the scattering matrix, is defined as the solution to the radial Schr\"{o}dinger equation that satisfies the boundary conditions \begin{equation} \label{gamowbound} \left\{ \begin{array}{rcl} \Omega_n(r) |_{r=0 \phantom{R}} & = & 0, \\ \frac{\mathrm{d}}{\mathrm{d}r}\, \Omega_n(r)|_{r=R \phantom{0}} & = & ik_n \Omega_n(r) |_{r=R}. \end{array} \right. \end{equation} The radius $R$ has to be chosen such that $R>R_0$, i.e., it is applied in the asymptotically free ($V=0$) region of the interaction potential.

The Gamow states behave asymptotically as $\Omega_n(r) \propto \exp[ik_nr]$. For the bound state poles $k_n = i\kappa_n$, with $\kappa_n$ a positive constant, the Gamow states are just the (properly normalized) bound state wave functions. However, for the poles with $\mathrm{Im}(k_n)<0$ the Gamow states exponentially diverge. Due to this exponential divergence, the Gamow states do not form an orthonormal basis for the $\mathcal{P}$ subspace. Defining a dual set of Gamow states as $\Omega_n^D \equiv \Omega_n^{\ast}$, the Gamow states do form a biorthogonal set in the sense that $\langle \Omega_n^D | \Omega_{n'} \rangle = \delta_{nn'}$. Here $\delta_{nn'}$ is the Kronecker delta, and the inner product is defined by means of analytic continuation in $k$ of the proper bound state eigensolutions to the resonance poles in the lower half of the complex $k$-plane \cite{Romo68,Kukulin89}. The normalization condition takes the form \begin{equation} \langle \Omega_n^D | \Omega_n \rangle = \int_0^R \Omega_n^2(r) \mathrm{d}r + \frac{i}{2k_n} \Omega_n^2(R) = 1. \end{equation} Note that, as before, $R$ has to be chosen such that $R>R_0$, in which case the normalization condition does not actually depend on the precise choice of $R$.

The Gamow state $|\Omega_n\rangle$ is an eigenstate of $H_{PP}$ and has an eigenvalue $E_n$. The dual state $|\Omega_n^D\rangle$ is an eigenstate of $H_{PP}^{\dag}$ with the eigenvalue $E_n^{\ast}$. This can be written in Dirac notation as \begin{eqnarray} H_{PP} |\Omega_n\rangle & = & E_n |\Omega_n\rangle ,\nonumber\\ \langle \Omega_n^D| H_{PP} & = & E_n \langle \Omega_n^D|. \end{eqnarray} From these last expressions it is immediately seen that the Hamiltonian $H_{PP}$ is diagonal with respect to the biorthogonal set of Gamow functions. The matrix elements of $H_{PP}$ can be evaluated as $\langle \Omega_n^D|H_{PP}|\Omega_{n'}\rangle = E_{n'} \delta_{nn'}$.

Using the Gamow resonance states, the resolvent operator $[E-H_{PP}]^{-1}$ can be expanded as a sum over eigenstates associated with poles of the scattering matrix. This type of expansion is known as the Mittag-Leffler expansion. The properties of the Mittag-Leffler expansion in terms of the Gamow resonance states have been extensively studied in the literature, see e.g. Ref.~\cite{Kukulin89,Garcia76,Romo78,Bang80}. We will therefore not give a derivation of the Mittag-Leffler expansion, but refer the reader to the literature. We here just give the Mittag-Leffler expansion of the resolvent operator: \begin{equation} \label{mittagleffler} \frac{1}{E-H_{PP}} = \sum_{n=1}^{\infty} \frac{| \Omega_{n} \rangle \langle \Omega_n^D |}{2k_n(k-k_n)}, \end{equation} where $n$ runs over all poles of the $S_P$ matrix. For notational convenience, we have used units such that $\hbar=2\mu=1$ so that $k$ corresponds to the energy as $E=k^2$. Note that $E$ can be any complex energy in Eq.~(\ref{mittagleffler}) in principle. In Sec.~\ref{SecInterplay} we discuss how the Mittag-Leffler expansion can be used to find the energy dependence of $A(E)$ analytically.

\section{Interplay between Feshbach resonances and potential resonances}\label{SecInterplay}

Inserting the Mittag-Leffler series Eq.~(\ref{mittagleffler}) in the complex energy shift Eq.~(\ref{AE}) results in \begin{equation} A(E) = \sum_{n=1}^{\infty} \frac{\langle \phi_b|H_{QP}| \Omega_{n} \rangle \langle \Omega_n^D |H_{PQ}|\phi_b \rangle}{2k_n(k-k_n)}. \end{equation} The complex energy shift of the unperturbed bound-state in the $Q$-channel is thus expanded over the contribution of the various poles of the $P$-channel scattering matrix. The exact expression for the complex energy shift derived here is the most important difference with previous work on the interplay between single-channel and multi-channel resonances.

We now illustrate the impact of a virtual state close to the collision threshold on the coupling between the spin channels $P$ and $Q$. The full $S$ matrix has the form $S = S_{\mathrm{bg}}^P S_{\mathrm{res}}^P S_{\mathrm{res}}^{\mathrm{Q}}$, and the poles of this $S$ matrix correspond to the energies of the dressed \mbox{(quasi-)}bound states. In Sec.~\ref{SecGamow} we already mentioned that in the case of $^{85}$Rb there is a virtual state close to threshold. We assume that this pole is dominant, and the effect of all other poles of the $P$-channel can be summarized by a background term. The energy shift due to coupling then takes the form \begin{equation} A(E) = \frac{ \langle \phi_B|H_{QP}|\Omega_{\mathrm{vs}} \rangle \langle \Omega_{\mathrm{vs}}^D|H_{PQ}|\phi_B \rangle }{ 2k_{\mathrm{vs}}(k-k_{\mathrm{vs}}) }. \label{Aapprox} \end{equation}

The basis $|\Omega_n\rangle$ is energy independent, as is the $Q$-channel bound state $|\phi_b\rangle$. The coupling $H_{PQ}=H_{QP}^{\dag}$ is energy independent as well. It then follows that the energy dependence of $A(E)$ is fully determined by the denominator in Eq.~(\ref{Aapprox}), which is proportional to $1/(k-k_{\mathrm{vs}})$.

As the numerator of Eq.~(\ref{Aapprox}) is a real constant and $k_{\mathrm{vs}}=-i\kappa_{\mathrm{vs}}$, we can write the complex energy shift as \begin{equation} A(E) = \frac{-iA_{\mathrm{vs}}}{2\kappa_{\mathrm{vs}}(k+i\kappa_{\mathrm{vs}})}, \end{equation} where $A_{\mathrm{vs}}=-\langle \phi_b|H_{QP}| \Omega_{\mathrm{vs}} \rangle \langle \Omega_{\mathrm{vs}}^D |H_{PQ}|\phi_b \rangle$ is a positive constant. The extra minus sign in this constant shows up for virtual states due to the normalization condition of the Gamow states.

For real and positive $k$ (or equivalently, for real and positive $E$ on the physical Riemann sheet) we can multiply the numerator and denominator with $(k-i\kappa_{\mathrm{vs}})$ and take the real and imaginary part of $A(E)$: \begin{eqnarray}  \Delta_{\mathrm{res}}(E) & = & \displaystyle \frac{-\frac{1}{2} A_{\mathrm{vs}}}{k^2+\kappa_{\mathrm{vs}}^2}, \\ \Gamma(E) & = & \displaystyle \frac{A_{\mathrm{vs}}k}{\kappa_{\mathrm{vs}} (k^2+\kappa_{\mathrm{vs}}^2)}. \label{Gammavirtual} \end{eqnarray} For $k=i\kappa$ on the positive imaginary axis (or equivalently, for real and negative $E$ on the physical Riemann sheet) the energy shift is real-valued and given as \begin{equation} \Delta_{\mathrm   {res}}(E) = \frac{-\frac{1}{2}A_{\mathrm{vs}}}{\kappa_{\mathrm{vs}} (\kappa+\kappa_{\mathrm{vs}})}. \end{equation}

The energies of the dressed states are given by the poles of the full $S$ matrix, and can be found by solving
\begin{equation}
(k+i\kappa_{\rm vs})(E- \epsilon_b(B) - A(E)) = 0.
\end{equation}
We see that in the presence of a virtual state pole close to threshold, the denominator of $\Delta_{\mathrm{res}}(E)$ gets close to zero for $E \rightarrow 0$. Therefore, the energy of the dressed molecular Feshbach state $\epsilon_{\mathrm{bind}} =\epsilon_b + \Delta_{\mathrm{res}}(E)$, strongly depends on the energy $E$ close to threshold. For positive energies, the energy shift $\Delta_{\mathrm{res}}(E)$ and width $\Gamma(E)$, related to the quasi-bound Feshbach state, depend strongly on the energy as well.

In order to give an accurate description of the dressed \mbox{(quasi-)}bound Feshbach state, it turns out that in the case of $^{85}$Rb the influence of the highest $P$-channel bound state on the energy shift and width should be taken into account as well (see Sec.~\ref{SecNumerical}). As can be seen in Fig.~\ref{CCbound}, this $P$-channel bound state and the $Q$-channel bound state have a broad avoided crossing. In the inset, the effect of the virtual state close to threshold is visible. However, on a larger scale it is clear that the \mbox{(quasi-)}bound state has still not converged to the bare $Q$-state energy.

The expressions for the energy shift and width including both pole contributions are given in the following. The pole associated with the highest $P$-channel bound state is denoted by $k_{\mathrm{bs}}=i\kappa_{\mathrm{bs}}$.
For real and positive $k$ (or equivalently, for real and positive $E$ on the physical Riemann sheet) we take the real and imaginary part of $A(E)$: \begin{eqnarray}  \Delta_{\mathrm{res}}(E) & = & \displaystyle \frac{-\frac{1}{2} A_{\mathrm{vs}}}{k^2+\kappa_{\mathrm{vs}}^2} + \frac{\frac{1}{2} A_{\mathrm{bs}}}{k^2+\kappa_{\mathrm{bs}}^2}, \\ \Gamma(E) & = & \displaystyle \frac{A_{\mathrm{vs}}k}{\kappa_{\mathrm{vs}} (k^2+\kappa_{\mathrm{vs}}^2)} + \frac{A_{\mathrm{bs}}k}{\kappa_{\mathrm{bs}} (k^2+\kappa_{\mathrm{bs}}^2)}. \label{Gammabound} \end{eqnarray} Here $A_{\mathrm{bs}}$ is a positive constant related to the coupling of the $P$-channel bound state to the $Q$-channel bound state. For $k=i\kappa$ on the positive imaginary axis (or equivalently, for real and negative $E$ on the physical Riemann sheet) the energy shift operator is real-valued and given as \begin{equation} \Delta_{\mathrm   {res}}(E) = \frac{-\frac{1}{2}A_{\mathrm{vs}}}{\kappa_{\mathrm{vs}} (\kappa+\kappa_{\mathrm{vs}})} + \frac{\frac{1}{2}A_{\mathrm{bs}}}{\kappa_{\mathrm{bs}} (\kappa+\kappa_{\mathrm{bs}})}. \end{equation}

The constants $A_{\mathrm{vs}}$ and $A_{\mathrm{bs}}$ can be found from input of coupled-channels calculations, or equivalently from measurement (see Sec.~\ref{SecNumerical}). Note that the contribution of the virtual state to the energy shift is negative valued, and the contribution of the proper bound state is positive valued.  Let us stress that there are no free parameters in this model. The model is fully characterized by only a few parameters, which are determined by physical quantities directly related to the true interaction potentials. These parameters can be extracted from numerical coupled-channels calculations, or directly from experimental measurements. More details are given in Sec.~\ref{SecNumerical}.

\section{Numerical method and results}\label{SecNumerical}

The numerical results in this section are based on coupled-channels calculations~\cite{coupledchannels} for rubidium, based on the most recent knowledge of the interaction potentials~\cite{Claussen03,Kempen02}. These calculations take the coupling between the relevant spin channels into account, i.e. $P$ and $Q$ are coupled. In order to find the physical properties of the open channel only, we `turn off' the coupling between $P$ and $Q$. This allows us to calculate the single-channel $S_P$ matrix, and find its poles and the background scattering length. In order to find the other model-parameters, we perform a calculation where the coupling between $P$ and $Q$ is again taken into account. In principle, only two data points are needed to determine the constants $A_{\rm vs}$ and $A_{\rm bs}$.

\begin{figure}
\includegraphics[width=\columnwidth]{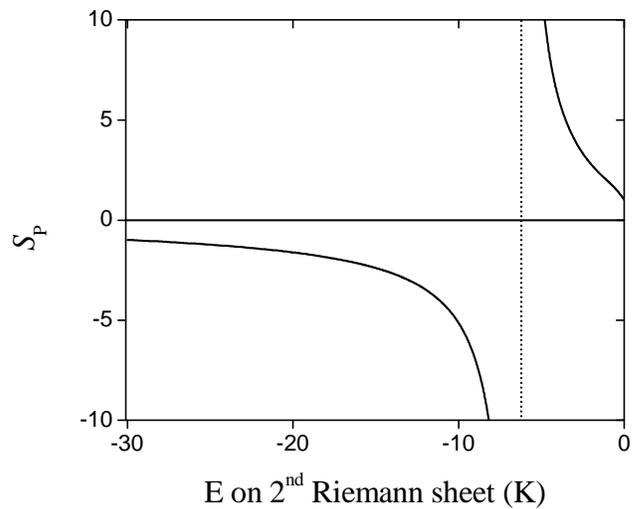}
\caption{$S_P$ matrix for negative energies on the non-physical Riemann sheet for $^{85}$Rb in the $|2,-2\rangle$ hyperfine channel. The pole of the $S_P$ matrix on this sheet is located at $\epsilon_{\mathrm{virtual}}/k_B = -6.45 \, \mu \mathrm{K}$.} \label{Sneg}
\end{figure}

In Fig.~\ref{Sneg} the $S_P$ matrix is shown for negative energies on the non-physical Riemann sheet, i.e. $\mathrm{Im}(k)<0$, for $^{85}$Rb in the $|2,-2\rangle$ hyperfine channel. The pole of the $S_P$ matrix on this sheet is located at $\epsilon_{\mathrm{virtual}}/k_B = -6.45 \, \mu \mathrm{K}$. This indicates that the $P$-channel interaction potential has a virtual state at this energy. The phase of the $S_P$ matrix for positive energies (on the physical sheet) was already shown in Fig.~\ref{Phase85Rb}.

In order to compare the results of our model with the numerical coupled-channels calculation, we perform a full calculation over a large energy and magnetic field range around threshold and resonance. For positive energies with respect to the $P$-channel threshold, the scattering phase $\delta(E,B)$ is calculated. At each energy value the derivative of $\delta(E,B)$ with respect to $B$ is calculated. According to Eq.~(\ref{smatrix}) the derivative is given by the following expression:
\begin{equation} \frac{\partial \delta(E,B)}{\partial B} =  \frac{ \Delta \mu^{\mathrm{mag}} \Gamma(E)/2 }{ [\epsilon_b^Q(B) + \Delta_{\mathrm{res}}(E) - E]^2 + [\Gamma(E)/2]^2 }.
\end{equation}
Here $\Delta \mu^{\mathrm{mag}}$ is the relative magnetic moment of the bare $Q$-channel bound state with respect to the $P$-channel threshold. This derivative function is the well-known Lorentz curve. The center of the Lorentz curve is given by the condition $E=\epsilon_b^Q(B) + \Delta_{\mathrm{res}}(E)$, which determines the location of the dressed quasi-bound Feshbach state. The energy width of this state equals the width of the Lorentz curve. We can thus calculate the position and width of the quasi-bound Feshbach state as a function of the collision energy.

For negative energies with respect to the $P$-channel threshold, the coupled-channels Schr\"{o}dinger equation is integrated outward, starting in the inner region of the interaction potential to some matching radius $r_m$. The Schr\"{o}dinger equation is also integrated inward from some $r_{\mathrm{max}}$ to the matching radius $r_m$. The boundary conditions at $r_{\mathrm{max}}$ are such that the solution asymptotically vanishes and is physically acceptable. If the two solutions obtained in this way are linearly dependent, the corresponding energy is an eigenenergy. At this energy, the full coupled potential has a bound state. This calculation is repeated as a function of magnetic field, and thus gives the field-dependent position of the dressed molecular Feshbach state.

\begin{figure}
\includegraphics[width=\columnwidth]{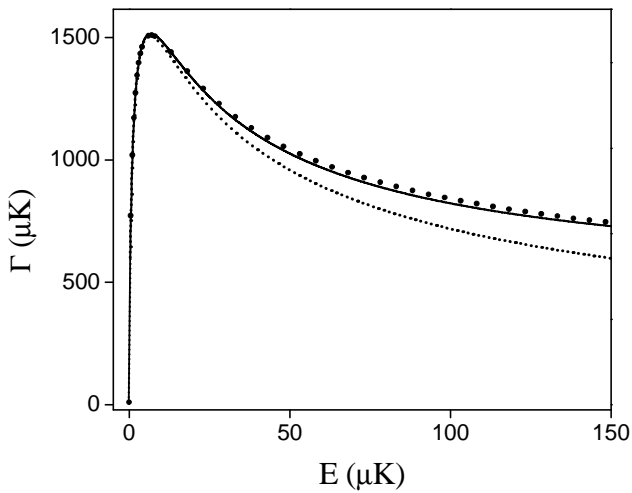}
\caption{Energy width of the dressed quasi-bound Feshbach state for $^{85}$Rb in the $|2,-2\rangle$ hyperfine channel. The dotted line shows the energy width according to Eq.~(\ref{Gammavirtual}), where only the virtual state, with $\kappa_{\mathrm{vs}} = 2.54\cdot 10^{-3} \, \mathrm{K}^{1/2}$ is taken into account, and $A_{\mathrm{vs}} = 1.94\cdot 10^{-8} \, \mathrm{K}^2$. The solid line shows the energy width according to Eq.~(\ref{Gammabound}), where the highest bound state of the $P$-channel is included as well, with $\kappa_{\mathrm{bs}} = 0.103 \, \mathrm{K}^{1/2}$. In this case $A_{\mathrm{vs}} = 1.92\cdot 10^{-8} \, \mathrm{K}^2$ and $A_{\mathrm{bs}} = 1.26\cdot 10^{-5} \, \mathrm{K}^2$.} \label{Width}
\end{figure}

The energy width of the dressed quasi-bound Feshbach state as calculated with a coupled-channels method is shown in Fig.~\ref{Width} (black dots). The dotted line shows the energy width according to our model, where only the $P$-channel virtual state is taken into account. The constant $A_{\mathrm{vs}}$ is determined by comparing Eq.~(\ref{Gammavirtual}) with a single low-energy data point. We see that for low energies the virtual-state expression and the coupled-channels data agree very well. However, for energies larger than roughly 10 microKelvin, the virtual-state expression starts to deviate.

The solid line shows the energy width according to our model, taking the highest bound state of the $P$-channel into account as well. The energy width is now accurately described by our model for a large energy domain. We obtain the two parameters $A_{\mathrm{vs}}$ and $A_{\mathrm{bs}}$ from a fit of Eq.~(\ref{Gammabound}) to two data points.

\begin{figure}
\includegraphics[width=\columnwidth]{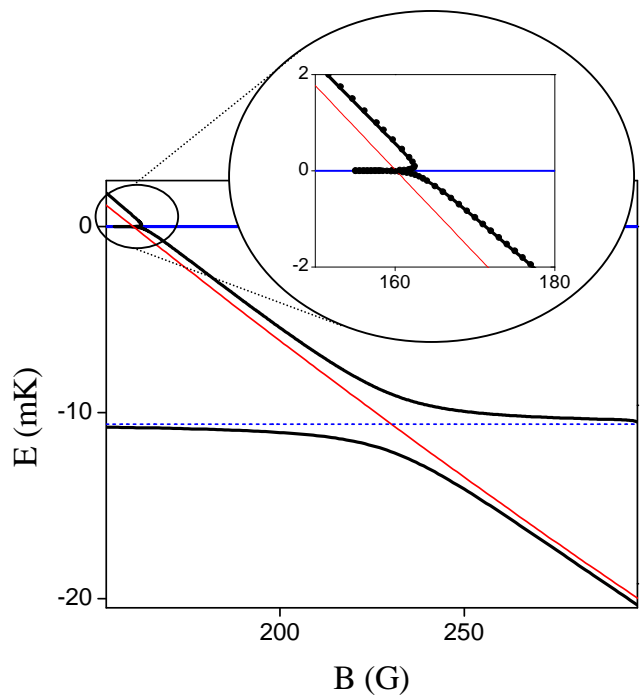}
\caption{(Color online) Energy of the dressed \mbox{(quasi-)}bound Feshbach state. The black dots indicate the coupled-channels data. The thick solid line indicates the energy according to our model. The thin solid line is the energy of the unperturbed $Q$-channel bound state, which around threshold is given by the linear expression $\epsilon_b^Q(B) = \Delta \mu^{\mathrm{mag}}(B-\overline{B}_0)$, with $\Delta \mu^{\mathrm{mag}} = -1.75\cdot 10^{-4} \, \mathrm{K/G} = -3.64 \, \mathrm{MHz/G}$ and $\overline{B}_0 = 160.1 \, \mathrm{G}$.} \label{CCbound}
\end{figure}

\begin{figure}
\includegraphics[width=\columnwidth]{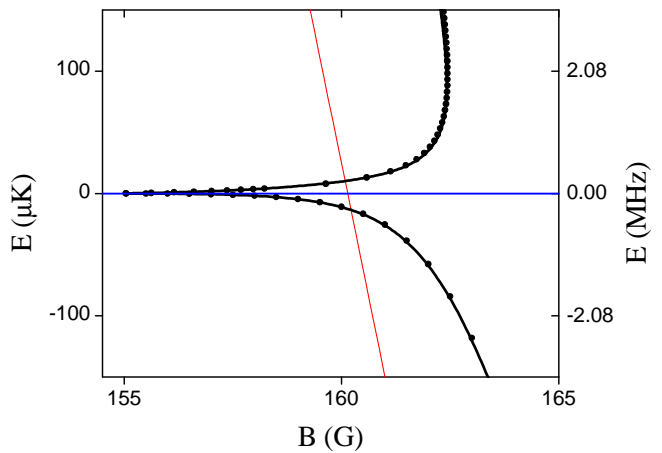}
\caption{(Color online) Same as in Fig.~\ref{CCbound}, but now for energies closer to the $P$-channel threshold.} \label{CCboundzoom}
\end{figure}

We now insert these parameters into the expressions for the energy shift $\Delta_{\mathrm{res}}(E)$ to describe the energy of the dressed \mbox{(quasi-)}bound Feshbach state. The result is shown in Fig.~\ref{CCbound}, where the black dots indicate the coupled-channels results and the solid line is obtained from our model. In Fig.~\ref{CCboundzoom} we zoom in closer to the $P$-channel threshold.

The unperturbed (or bare) bound state in the $Q$-channel subspace is dressed by the coupling to the $P$-channel subspace. This induces an avoided crossing with the highest $P$-channel bound state. The avoided crossing is broad in the sense that, even though the unperturbed $P$-channel bound state is located at roughly $\epsilon_{\mathrm{bound}}/k_B \approx -10 \, m\mathrm{K}$, close to the $P$-channel threshold the dressed state still has not converged to the bare $Q$-channel bound state.

The $P$-channel virtual state is not located at the physical energy sheet, and there is no avoided crossing of the usual kind between the dressed $Q$-channel \mbox{(quasi-)}bound state and the virtual state. However, the virtual state is located close to the collision threshold and induces a strong threshold effect. This threshold effect dominates the behavior of the molecular binding energy near the collision threshold, and has to be taken into account explicitly. In our model we take the relevant $P$-channel bound and/or virtual states into account analytically. From these figures it is immediately seen that our model agrees perfectly with full coupled-channels calculations for a very large energy domain. The binding energy of the dressed molecular state that has been measured in Ref.~\cite{Donley02,Claussen03} is described analytically with high precision.

\section{Other approaches} \label{SecOther}

\begin{figure}
\includegraphics[width=\columnwidth]{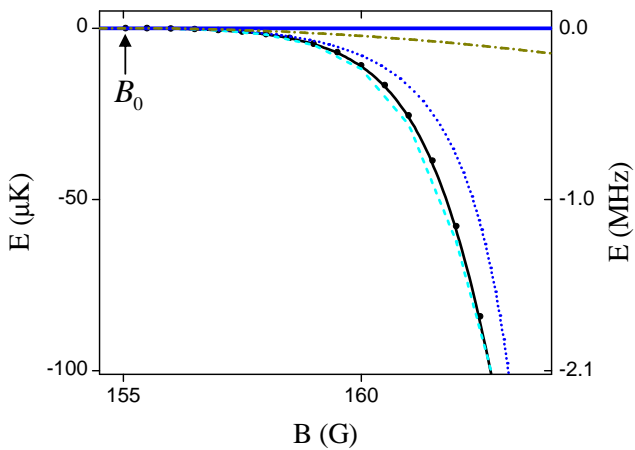}
\caption{(Color online) Comparison of coupled-channels calculations of the binding energy (black dots) with the contact model (dotted line), the single-resonance Feshbach model (dashed-dotted line), the effective range model (dashed line), and our virtual state model (solid line). Although for the magnetic fields shown the effective range model and our model give comparable results, our model gives an analytical expression for the binding energy. This is not the case for the effective range model, where a field-dependent parameter $r_0(B)$ has to be taken into account that has to be calculated numerically.} \label{Other}
\end{figure}

In this section, we compare our model with some other approaches commonly used in the description of Feshbach resonances. A model which is conveniently used in many-body theories is the contact potential (or zero range potential). In this approach the real interaction potentials are replaced by deltafunctions or pseudopotentials, proportional to the s-wave scattering length $a$. In the vicinity of a Feshbach resonance the dispersive formula Eq.~(\ref{aofB}) is used. The scattering matrix takes the form \cite{Newton82}
\begin{equation}
S(k)=\frac{1-ika(B)}{1+ika(B)}.
\end{equation}
The molecular binding energy is determined by the pole of $S(k)$, and is given as $\epsilon_{\mathrm{bind}}=-\hbar^2/(2\mu a^2(B))$. In Fig.~\ref{Other} we compare the resulting binding energy (dotted line) with the coupled-channels results (black dots). It is clear that close to the resonance magnetic field $B_0$, the contact model binding energy agrees quite well with the coupled-channels binding energy. Further away from the resonance, the contact model energy starts to deviate significantly from the exact binding energy.

In Eq.~(\ref{singleresonance}) the $S$ matrix for the single-resonance Feshbach model is given. The pole of this $S$ matrix gives the energies of the dressed \mbox{(quasi-)}bound state. This energy is indicated by the dashed-dotted line in Fig.~\ref{Other}. It can be clearly seen, as expected, that the single-resonance model fails already relatively close to threshold.

Another commonly used model originates from the effective range approach. In the contact model the scattering phase $\delta(k)$ is approximated as $\delta(k)=-\arctan[ka]$, but this approximation already breaks down at low energies. In the effective range model a second term is included to describe $\delta(k)$ further away from the collision threshold. The scattering matrix is given as
\begin{equation}
S(k) = \frac{-1/a(B) + r_0(B) k^2/2 +ik}{-1/a(B) + r_0(B) k^2/2 -ik},
\end{equation}
where $r_0(B)$ is the effective range parameter~\cite{Newton82}. The molecular binding energy can again be found by solving \begin{equation} -\frac{1}{a(B)} + \frac{1}{2} r_0(B) k^2 -ik =0, \end{equation} and is shown in Fig.~\ref{Other} (dashed line). We determined $r_0(B)$ and $a(B)$ simultaneously from a coupled-channels calculation, by fitting to the scattering phase for two different energy values, for several values of the magnetic field. Although the field-dependent effective range model agrees reasonably well with the coupled-channels binding energy for magnetic fields close to $B_0$, it breaks down for fields larger than $B\approx 164$ Gauss and cannot be applied further away from threshold.

Summarizing, the contact model, single-resonance Feshbach model, and effective range model are low-energy approximations of the exact scattering matrix. These approaches give reasonable agreement with the coupled-channels molecular binding energy close to the resonant magnetic field $B_0$, but for energies further away from threshold these descriptions give poor agreement.

Our model has several important advantages compared to the models discussed here. First of all, our model is directly based on the underlying physics of the interplay between potential and Feshbach resonances. The contact potential model is based on the assumption that the scattering phase can be replaced by its low-energy limit, $\delta(k) = -\arctan[ka(B)]$. As we have seen, the potential resonance and Feshbach resonance introduce additional energy-dependence of the scattering phase shift. The single-resonance Feshbach model would give an adequate description for the situation where $a_{\rm bg}$ is on the order of the range of the potential. However, when $a_{\rm bg}$ is large compared to this range, this model breaks down as well. The effective range approach gives a better description of the binding energy compared to the two previous models. However, it still breaks down at some point, and moreover, it does not give much physical insight in the mechanism behind the additional energy dependence of various cold-collision properties. Another clear disadvantage of the effective range approach is that we have to calculate $r_0(B)$ numerically as a function of the magnetic field using a coupled-channels method. In our model the binding energy is given by a simple analytical formula, which gives an excellent description even for magnetic fields far from $B_0$ and energies of order milliKelvin. Our model is fully characterized by only a few parameters, which can be extracted from coupled-channels calculations, or directly from measurements.

\subsection*{Two resonances in $\mathcal{Q}$-space}

Another approach that has been proposed in Ref.~\cite{Kokkelmans02} is to use a double-resonance parametrization of the scattering matrix within the Feshbach projection formalism. The second resonance introduced should account for the influence of the potential resonance on the properties of the Feshbach resonance.

The Feshbach projection formalism can be used to describe potential resonances, albeit in a rather indirect manner. In a paper by Domcke \cite{Domcke83}, the Feshbach projection-operator approach is used to describe potential resonances in scattering systems. The projector onto the resonance states is defined as $Q = \sum_{n=1}^N |\phi_n\rangle \langle \phi_n|$, where the set of functions $\{\langle r|\phi_n\rangle\}$ is an arbitrary orthonormal set of square-integrable functions. The formal requirement on the projection operator $Q$ is given as \begin{equation} \label{limitQ} \langle r|Q|\psi \rangle \: \begin{subarray}{c}\phantom{=} \\ \displaystyle \rightarrow \\ \scriptstyle{r \rightarrow \infty} \end{subarray} \: 0 \qquad \textrm{for any } |\psi \rangle. \end{equation} The formalism leads to a decomposition of the scattering matrix into a resonant and a non-resonant part, where the resonances are all contained in $\mathcal{Q}$ space if $N$ is chosen large enough. The choice of the states $|\phi_n\rangle$ to be used is arbitrary, however, and the correspondence between these states and the real resonance states (such as the Gamow states) is not clear. More specifically, the Gamow states in general do not satisfy relation Eq.~(\ref{limitQ}) and cannot be used to construct an operator onto the real resonance states within this formalism.

However, as Domcke argues, increasing the number of states used in the projector $Q$ by one, seems to remove exactly one resonance pole from the background part of the scattering matrix. One could therefore hope that adding an appropriate bound state $|\phi_i\rangle$, the virtual state resonance due to the Gamow resonance state $|\Omega_{\mathrm{virtual}}\rangle$ can be indirectly included in the $\mathcal{Q}$ subspace. This is the approach followed in Ref.~\cite{Kokkelmans02}, in order to account for the interplay between a Feshbach resonance and a potential resonance in $^6$Li collisions. We will refer to this as the double-resonance approach.

Although the bound state introduced to account for the second resonance is not clearly linked to the $P$-channel virtual state that gives rise to the potential resonance, the double-resonance approach is mathematically equivalent to our model under some constraints. We will show this in the following.

Introducing two $Q$-channel bound states $|\phi_i \rangle$, one to account for the Feshbach resonance ($i=b$) and one to account for the virtual state resonance ($i=v$), the direct part of the scattering matrix does not contain a resonant feature anymore and takes the form $S_{\mathrm{bg}}^P=\exp[-2ika^P_{\mathrm{bg}}]$, where $a^P_{\mathrm{bg}}$ is on the order of the range of the interaction potential. The $\mathcal{Q}$ subspace contains the two resonances introduced above, and the total scattering matrix takes the form \cite{Kokkelmans02}
\begin{equation}
S(k) = S_{\mathrm{bg}}^P \left(1- \frac{2ik (C_b \Lambda_v + C_v \Lambda_b) }{ \Lambda_b \Lambda_v +ik(C_b \Lambda_v + C_v\Lambda_b) }  \right).
\end{equation}
Here $\Lambda_i=E - \epsilon_i^Q - \Delta_i$ is the detuning of the dressed state $i$ with energy $\epsilon_i^Q + \Delta_i$. The decay width of the dressed state $i$ is given by $\Gamma_i=2C_ik$. Note that the energy shift of both resonance states is approximately constant, and the decay width scales with $k$ according to the Wigner threshold law. This is a direct consequence of the removal of the low-energy resonance from the direct channel by including the state $|\phi_v \rangle$ into the $\mathcal{Q}$ subspace.

The poles of the scattering matrix determine the (complex) energy of the dressed resonance states. In the double-resonance approach these poles are determined by the condition
\begin{equation}
\Lambda_b \Lambda_v +ik(C_b \Lambda_v + C_v \Lambda_b)=0.
\end{equation}
In our model the poles are determined by the condition
\begin{equation}
(k+i\kappa_{\mathrm{vs}}) (E - \epsilon_b^Q - A(E))=0,
\end{equation}
if we only take the $P$-channel virtual state and the $Q$-channel bound state into account. In other words, we neglect the avoided crossing with the highest $P$-channel bound state. Under the constraint that the dressed bound state that removes the potential resonance from the direct channel is located far away from the collision threshold, i.e., $\Lambda_v \simeq -\epsilon_v - \Delta_v$, the two approaches are equivalent if the following relations are satisfied:
\begin{eqnarray}
\frac{ \Lambda_v }{ C_v } &=& -\kappa_{\mathrm{vs}} ,\\
C_b &=& -\frac{ \Delta_b }{ \kappa_{\mathrm{vs}} }.
\end{eqnarray}
These relations show how the double-resonance parameters are related to the position of the virtual state (described by $\kappa_{\mathrm{vs}}$), and the zero-energy shift of the dressed Feshbach resonance state, $\Delta_{\mathrm{b}}=\Delta_{\mathrm{res}}(E=0)$. Note that the energy $E$ always has to be negligible compared to $\epsilon_v+\Delta_v$, otherwise the double-resonance S-matrix will introduce a non-physical energy-dependence in the scattering phase-shift.

The double-resonance model does give an equivalent description of the scattering process, and can be used to parameterize the scattering matrix and/or scattering length. The link between the properties of the open-channel resonance and the second resonance introduced in the double-resonance model is not really clear. Our model has the important advantage that it is directly related to the underlying physics giving rise to the $P$-channel resonance.

\section{Conclusions} \label{SecConclusions}

In this paper, we have derived an analytical model that describes the cold-collision properties of two interacting particles near a Feshbach resonance with large background scattering length. The large background scattering length results from an open-channel resonance near threshold, and this resonance has to be treated explicitly. The open-channel scattering is included in the Feshbach theory of resonances via a contribution from its poles of the $S$ matrix, and its non-resonant open-channel background scattering length. Here the latter corresponds to the true range of the potential. As an example, we study the $B_0 = 155$~G Feshbach resonance of $^{85}$Rb, and we show that our model compares excellent with numerical coupled-channels calculations in a large range of energies around threshold.

Our model offers a simple physical picture for the understanding of threshold effects around resonance. It explains how the energy-shift between the dressed \mbox{(quasi-)}bound Feshbach state and the bare molecular state can be found.
In the literature, several other models have been proposed to describe the formation of (quasi-bound) molecules (see for instance~\cite{timmermans,greene,abeelen,mies}). We discuss in the following how our model compares to some other models.

Several ideas have been proposed to describe the formation of (quasi-bound) molecules. In some descriptions (c.f.~\cite{timmermans,abeelen,mies}) the threshold and (bare) molecular state are regarded as a two-level system, with an effective coupling that accounts for the coupling between the open and closed channels. This results in a Landau-Zener crossing between the threshold state and the molecular state. As can be seen from Fig.~\ref{Avoided_virtual}, a Landau-Zener crossing does not correctly describe the dressed molecular state in case the open channel has a virtual state near threshold. The difference is striking, since the quasi-bound state in the continuum starts from threshold at $B_0$ (where $a$ is infinity), and then curves towards the bare Feshbach state. In the Landau-Zener model the quasi-bound state curves towards threshold at the other side of the bare Feshbach state. In case there is a real bound state in the open channel, the behavior of the dressed quasi-bound state appears to be qualitatively in agreement with the Landau-Zener crossing (c.f.~Fig.~\ref{Avoided_bound}). However, this is only the case since there is a real avoided crossing below threshold between the Feshbach state and the open-channel bound state. Threshold effects (visible in the inset) are not accounted for in the Landau-zener crossing.

In other descriptions (c.f.~\cite{greene}) the energy levels of an interacting pair of atoms in a harmonic trap are calculated. Here the lowest state behaves as the Feshbach bound state in the free-particle case for $a$ small and positive, and connects to the lowest harmonic oscillator state for $a$ small and negative. All other trap states start on one side of the resonance ($a>0$) in a particular trap state $n$, and are connected on the other side of the resonance ($a<0$) to a trap state $n+1$. There is a strong field dependence of the eigenenergies for field values close to resonance, and the bare trap-levels can be regarded as having avoided crossings with the quasi-bound state. This can be understood from analytical models of two trapped atoms~\cite{fumi,kokkelmans,pricoupenko}, which calculate the energy shift of the trapped atoms due to interactions, using the scattering phase as a boundary condition for the trap-wavefunction of the particles. In this way, it can be seen that the level-energy is strongly field dependent when it coincides with the quasi-bound Feshbach state in the free-atom case.
In the situation of a virtual state in the open channel (c.f.~Fig.\ref{Avoided_virtual}), the quasi-bound state energy changes its slope in a region close to threshold. The avoided crossings between bare trap-levels and quasi-bound Feshbach state should therefore follow this behavior.

The interplay between an open-channel resonance and a Feshbach resonance is usually not taken into account in the description of resonance many-body systems~\cite{kokkelmans2,mackie,kohler,combescot,yurovski,stoof}. The open-channel will have an important effect in a regime where the energies of the system are of order of or larger than the energy of the open-channel virtual state. A clear example where this effect is visible is the Ramsey-fringe experiment with $^{85}$Rb~\cite{Donley02,Claussen03}, where the range of binding energies is so large that the bound state cannot be properly described by a single-resonance model.

\section{Acknowledgements}
We thank H.~C.~W.~Beijerinck for stimulating discussions. This work was supported by the Netherlands Organisation for Scientific Research (NWO). S.~J.~J.~M.~F.~K.~acknowledges also a Marie Curie grant from the E.U. under contract No. MCFI-2002-00968. E.~G.~M.~v.~K.~acknowledges support from the Stichting FOM, which is financially supported by NWO. Laboratoire Kastler Brossel is a {\it Unit\'e de Recherche de l'Ecole Normale Sup\'erieure et de l'Universit\'e Paris 6, associ\'ee au CNRS.}


\begin{thebibliography}{100}
\bibitem{notepoles} Note that virtual-state poles can collide on the negative imaginary $k$-axis, and move into the third and fourth quadrant of the complex $k$ plane, see e.g. \cite{Nussenzveig59}. However, these poles are located sufficiently far from the real $k$-axis, such that they can be included in the background part of the scattering matrix.
\bibitem{notedipolar} Note that similar behavior was found in the case of dipole-modified s-wave scattering of low-energy electrons from weakly dipolar molecules \cite{Domcke81}. However, the physical mechanism responsible for this behavior is different, and the open channel studied in this reference does not contain a potential resonance. Moreover, in contrast to our approach, the work in this reference is based on a phenomenological parametrization of the decay width $\Gamma(E)$.
\bibitem{notefiniterange} Note that $R_0$ can be taken very large, and that for all practical purposes the difference between the finite-range potential and the real potential are negligible in this case \cite{Taylor72}. Mathematically, for potentials which behave asymptotically as $1/r^n$ and whithout a cutoff, the Jost function has a branch point at the origin, see e.g. \cite{Domcke81,Herzenberg84}. Within the finite-range approximation the Jost function is an entire function of $k$.
\end{thebibliography}
\end{document}